\documentclass{article}

\usepackage{amsmath}
\usepackage{multirow}
\usepackage{float}

\usepackage{authblk}
\usepackage{hyperref}
\usepackage{pdfpages}
\usepackage{appendix}

\usepackage{graphicx}

\newcommand*{\email}[1]{%
    \normalsize\href{mailto:#1}{#1}\par
    }
\title{Examining Correlation Between Trust and Transparency with Explainable Artificial Intelligence}
\author{Arnav Kartikeya}
\affil{\email{arnavkartikeya@gmail.com}}

\begin{document}

\maketitle

\begin{abstract}
Trust between humans and artificial intelligence(AI) is an issue which has implications in many fields of human computer interaction. The current issue with artificial intelligence is a lack of transparency into its decision making, and literature shows that increasing transparency increases trust. Explainable artificial intelligence has the ability to increase transparency of AI, which could potentially increase trust for humans. This paper attempts to use the task of predicting yelp review star ratings with assistance from an explainable and non explainable artificial intelligence to see if trust is increased with increased transparency. Results show that for these tasks, explainable artificial intelligence provided significant increase in trust as a measure of influence. 
\end{abstract}

\maketitle

\section{Introduction}

Trust in automation, primarily artificial intelligence, is an important issue in the field of human computer interaction. Prior literature has shown that with an increase in transparency, trust increases as well \cite{article}. Artificial Intelligence(AI) has the well known problem of a lack of justification of its decision making, therefore decreasing transparency between a user and the AI. Explainable artificial intelligence(XAI) provides increased transparency to the user, which has the potential of increasing trust. The XAI used in this paper is LIME, an algorithm developed in 2016 for model agnostic explanations into model decision making \cite{lime}. This paper attempts to find the correlation between XAI, increased transparency, and trust in specific tasks. 

A common task for artificial intelligence is to suggest or aid a human in a task. Recommendation algorithms used by Netflix, or systems aiding pilots in spatially disorienting situations are prime examples of this task. Literature examining the relation between trust and transparency in these tasks exist, as well as literature on the impact of XAI on human interaction \cite{paper}. Our work makes the following contributions: It examines trust through the lens of XAI's influence of human decision making, and uses metrics of quantitative data as oppose to subjective opinion-based surveys. Influence in this case is defined as how often a user changes their decision with additional aid from an artificial intelligence model. Rather than using metrics of opinion-based surveys, this paper examines how exactly the model's decision impacts human decisions, which will show how trust and transparency correlate in realistic scenarios rather than subjective opinions. I also examine the effectiveness of opinion-based surveys such as the Trust in Automation Questionnaire in how it represents actual influence of a model in a realistic scenario.

I hypothesize the following:

\begin{enumerate}
\item \label{Perf1}the LIME XAI model will influence human decisions significantly more than a normal artificial intelligence, \item \label{Perf2} trust measured through the Trust in Automation opinion-based survey \cite{airforce} will correspond with the actual influence from artificial intelligence.
\end{enumerate}

\section{Experimental Design}

In order to experimentally examine the relation between trust and transparency in the task described previously, a method of two separate surveys given to two separate groups was used. The surveys both asked the respondent to complete the task of predicting the star rating an individual gave to a restaurant with only the information of the text in the review, and a machine learning model's output. The surveys only vary in the amount of information given by the model, meaning the exact texts given and questions asked were the exact same. The following section elaborates on the differences and similarities. The machine learning model in both cases were the same, they both were pre-trained facebook fasttext models \cite{fasttext}, and both were trained on the same official yelp dataset. 
\subsection{Similarities between surveys}

Both surveys follow the same two sequence set of questions, and these sequences repeat 15 times to create 30 questions. The sequence is as follows. The first question in the sequence asks the user to predict, on a scale of 1 to 5 stars, what they believe the Yelp review associated with the question rated the restaurant. The second question gives a model output in the form of an image, which is either the label of the class of yelp review (in the form of \emph{\_\_label\_\_X.0}, where X represents the number 1 through 5 for stars). Depending on the survey, either explanation is shown. After all the 15 two-sequence questions are answered by the respondent, they will answer the Trust In Automation Questionnaire on a 1 to 7 likert scale.

\subsection{Differences between surveys}

The previous section mentions that either class label or Figure 1 is used as the model’s output, and this depends on the survey being answered. The first survey, known as the basic survey, only offers class label, which provides no transparency into the model’s decision making.  

The second survey, known as advanced survey, uses Figure 1 as its output. From left to right, the first section shows the confidence that the model had in its classification of each star. The figure shows that it had a confidence of 0.03 for 5 stars. The second portion shows the top features, or words, that influenced the model’s decision of the label it chose and the ones it did not. A higher number represents a larger influence in its decision. The third section repeats the text from the Yelp review, and highlights those specific words that were mentioned in the second section. A darker highlight corresponds to higher influence. This figure shows greater transparency in the model’s decision than Figure 1 does.

The explanation for Figure 1 was created by LIME. In both instances, a pretrained Fast-text natural language processing model was further trained on the official Yelp dataset to accomplish the task of predicting the star rating of a review.

\begin{figure}[H] \label{fig1}
    \centering
    \includegraphics[scale=0.45]{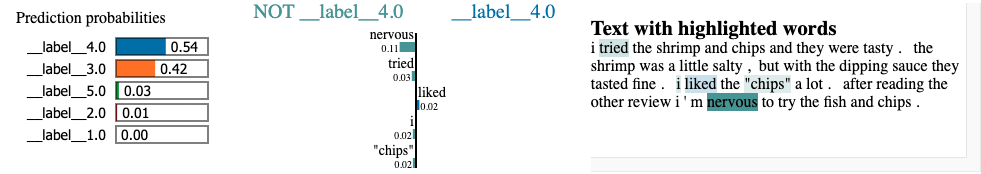}
    \caption{Explanation provided for second survey. (Left) Confidence in each class prediction, (Middle) Top most influential words in decision making, (Right) Top influential words, highlighted within entire text}
\end{figure}

\subsection{Distribution of survey}

Both surveys were separate tasks distributed on Amazon Mechanical Turks. 25 participants were used for each survey, with no additional criteria included. The sets of participants were not the same set of individuals, and were anonymous.

\section{Evaluation of survey results}

In order to quantify the 50 respondent answers and understand the relation between trust and transparency, data was split into multiple groups and compared between the two surveys using customized metrics. 

\subsection{Splitting the data}

Data was split by first which survey was being answered. The first group contained answers from the basic survey, and the second from the advanced. In each group, further subgroups were created. In both surveys, the model was correct in its prediction 6 times, 6 were incorrect by 1 or 2 stars, and 3 by more than 3 stars. I compared these groups between the surveys in order to ensure confidence was not the leading factor in decision making, and to determine exactly how much transparency can influence a users decision's. Measuring if there was difference in the sets of data for all three would ensure that transparency increases trust in all cases, however if there is not a difference in the less correct sets, it would show that confidence has a greater influence on trust and transparency cannot overcome the lack of confidence in the model.

\subsection{Comparing the data}

To compare the data between the groups, a custom metrics was created, the model influence metric. This was done to show the exactly how the model changed the users decision. A metric of checking if the user changed their answer between originally answering and being presented with the model's decision would not suffice, because there is the likelihood the respondent further lowered their answer away from the model's decision, as they simply rethought their answer. Therefore, a metric was needed which checked if the respondent changes their answer to be closer to the model. This would show that the respondent did use the model's output in their decision making, showing trust between the model and the respondent. If the respondent’s answer was closer to the model’s output by the second question, then a score for the respondent was incremented by 1. For example, if a respondent believed yelp review to have a star prediction of 2 stars, but the model predicted 4 and the respondent then changed their answer to 3, that would increment their score by 1. This was done for each of the subsections based on correctness, and the results were compared between the basic and advanced sets of data as mentioned previously. One case which presented an issue was the case of when the respondent guessed the same answer as the model did, without the model's input. If they guessed 3 stars before looking at the model, an the model also guessed 3 stars, then no trust is being measured if the respondent keeps their answer. To prevent this from occurring, I omitted those answers from the data and kept the metric as a percentage value. This ensured that those data points would not be included, while still maintaining a proper metric that can be compared over varying sizes of data. The following section will discuss the results. Equation 1 shows the formula for the metric used.

\begin{equation} 
    \scalebox{2}{$\left ( \frac{100\%}{25}  \right ) (\sum \frac{\text{\#influenced answers}}{\text{\#non omitted answers}}) $} 
    \label{eq1}
\end{equation}

\section{Results}

\subsection{Model influence metric results}
After processing all 50 survey responses using metrics and groups described in the previous section, the results were created and displayed in the Figures and tables below. Between the groups of correct, slightly incorrect, and totally incorrect, the metric used showed a statistically significant difference. A one tailed independent mean t-test was used for finding the p values listed in the table. In all cases p value is less than 0.05. Table 1 shows the individual means of all 6 groups compared. 

\subsection{Trust evaluation results}

Questions taken from the trust in automation questionnaire were placed at the end of both surveys and analyzed in order to see if there exists any contrast between the subjective trust measured by the questionnaire and the metric used in this paper. Both group's surveys were averaged. The advanced questionnaire had an average of 4.50 and the basic an average of 4.71 (p-value=0.14, one-tailed independent mean t-test).The results show no statistically significant difference between the advanced and basic surveys, which does not align with the difference shown in the previous section with the model influence metric.

\begin{table} \label{table}
\centering
\begin{tabular}{ |c|c|c|c| } 
\hline
{Groups} & {Basic Survey} & {Advanced Survey} & {p-value} \\
\hline
{Correct} & {33.27\%} & {59.54\%} & {.003} \\ 
{Slightly In.}& {33.20\%} & {55.67\%} & {.009} \\ 
{Totally In.}& {50.00\%} & {65.34\%} & {.038}\\ 
\hline
\end{tabular}
\caption{Difference between basic and advanced survey in means, p-value calculated by one-tailed t-test}
\end{table}

\subsection{Discussion}
The purpose of this study was to experimentally verify if transparency increases trust in the case of artificial intelligence aided tasks similar to the one described. Results from the experiment show statistically significant difference between the sets of data, and therefore do not reject hypothesis 1. This shows that with increased transparency, trust increases as well between the respondent and the model, at least when trust is measured as model's influence on the respondent. The data also shows that any additional insight into the model's decision making will increase trust, regardless of if the model is totally incorrect in its results. This is shown by comparing the results from the "totally incorrect" sections of the transparent and non transparent models. 

The second section of the experiment pertained to the Trust in Automation questionnaire, and whether its results align with that from the surveys. The confidence and dependence questions which were discussed in previous sections show no statistically significant results between transparent and non transparent models, which is not reflected by the metrics used for this paper. This shows that for measuring model influence through change in decision making after model input, the Trust in Automation questionnaire does not align with measures used. Subjective trust measured through the Trust in Automation Questionnaire therefore contrasts with trust measured through the metric used in this paper in experimental scenarios. This paper comes to its conclusions for the specific experiment and metric used in the paper. This paper serves as a starting ground for future research in verifying the hypothesis mentioned for different experiments and metrics, as well as larger sets of data. 

\bibliographystyle{plain}
\bibliography{main}

\begin{thebibliography}{1}

\bibitem{article}
Kevin Hoff and Masooda Bashir.
\newblock Trust in automation: Integrating empirical evidence on factors that
  influence trust.
\newblock {\em Human Factors The Journal of the Human Factors and Ergonomics
  Society}, 57:407--434, 05 2015.

\bibitem{airforce}
Jiun-Yin Jian, Ann~M. Bisantz, and Colin~G. Drury.
\newblock Foundations for an empirically determined scale of trust in automated
  systems.
\newblock {\em International Journal of Cognitive Ergonomics}, 4(1):53--71,
  2000.

\bibitem{fasttext}
Armand Joulin, Edouard Grave, Piotr Bojanowski, and Tomas Mikolov.
\newblock Bag of tricks for efficient text classification.
\newblock {\em arXiv preprint arXiv:1607.01759}, 1, 2016.

\bibitem{lime}
Marco~T{\'{u}}lio Ribeiro, Sameer Singh, and Carlos Guestrin.
\newblock "why should {I} trust you?": Explaining the predictions of any
  classifier.
\newblock {\em CoRR}, abs/1602.04938, 2016.

\bibitem{paper}
Katharina Weitz, Dominik Schiller, Ruben Schlagowski, Tobias Huber, and
  Elisabeth Andr\'{e}.
\newblock "do you trust me?": Increasing user-trust by integrating virtual
  agents in explainable ai interaction design.
\newblock {\em Proceedings of the 19th ACM International Conference on
  Intelligent Virtual Agents}, IVA '19:1--3, 07 2019.

\end{thebibliography}

\end{document}


\appendix
\appendixpage
\addappheadtotoc

\tableofcontents

\section{Advanced questionnaire}

\includepdf[page={1,2,3,4,5,6,7,8,9,10,11,12,13,14,15,16,17,18,19,20,21,22,23,24,25,26,27,28,29,30,31,32,33}]{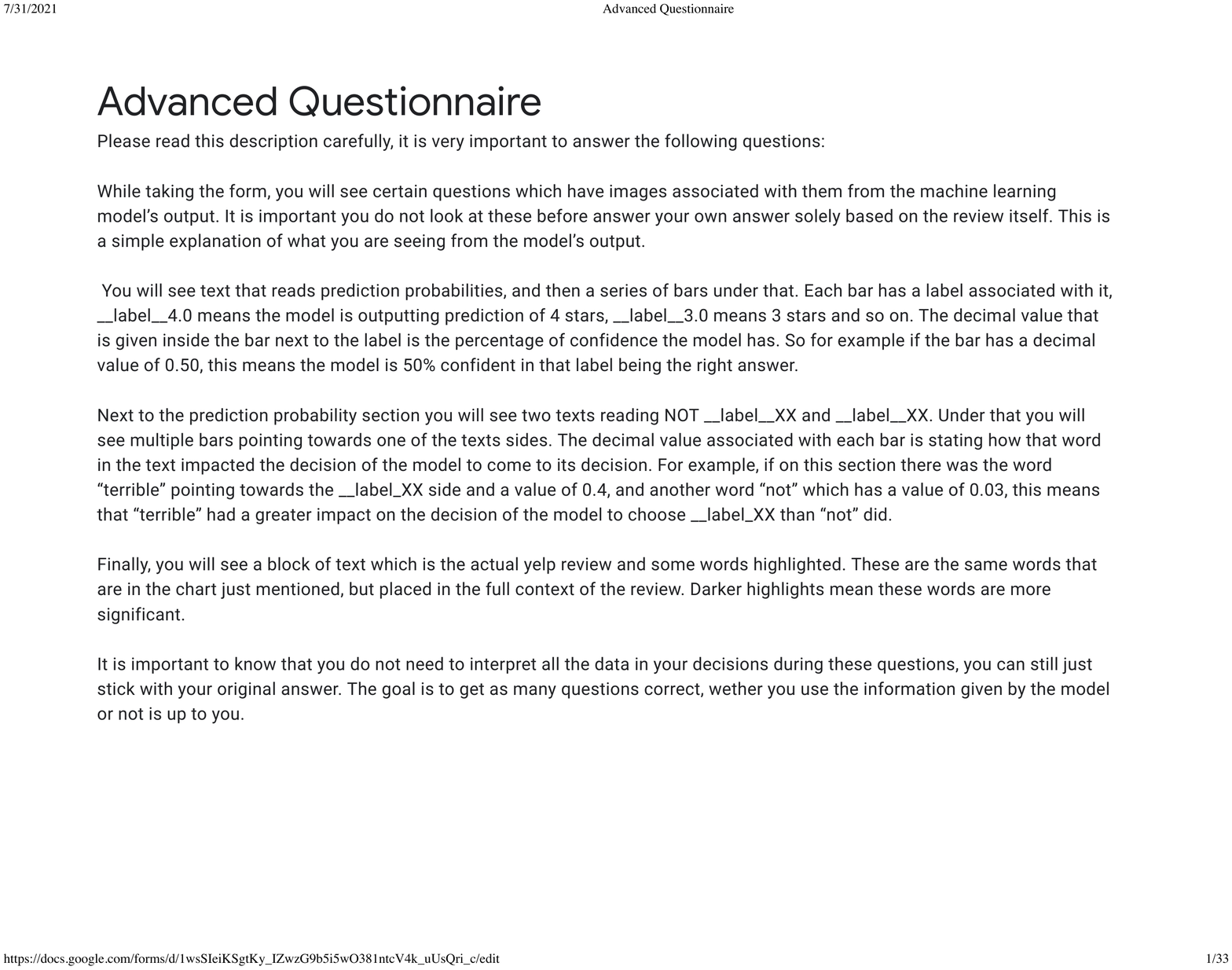}

\section{Basic questionnaire}

\includepdf[page={1,2,3,4,5,6,7,8,9,10,11,12,13,14,15,16,17,18,19,20,21,22}]{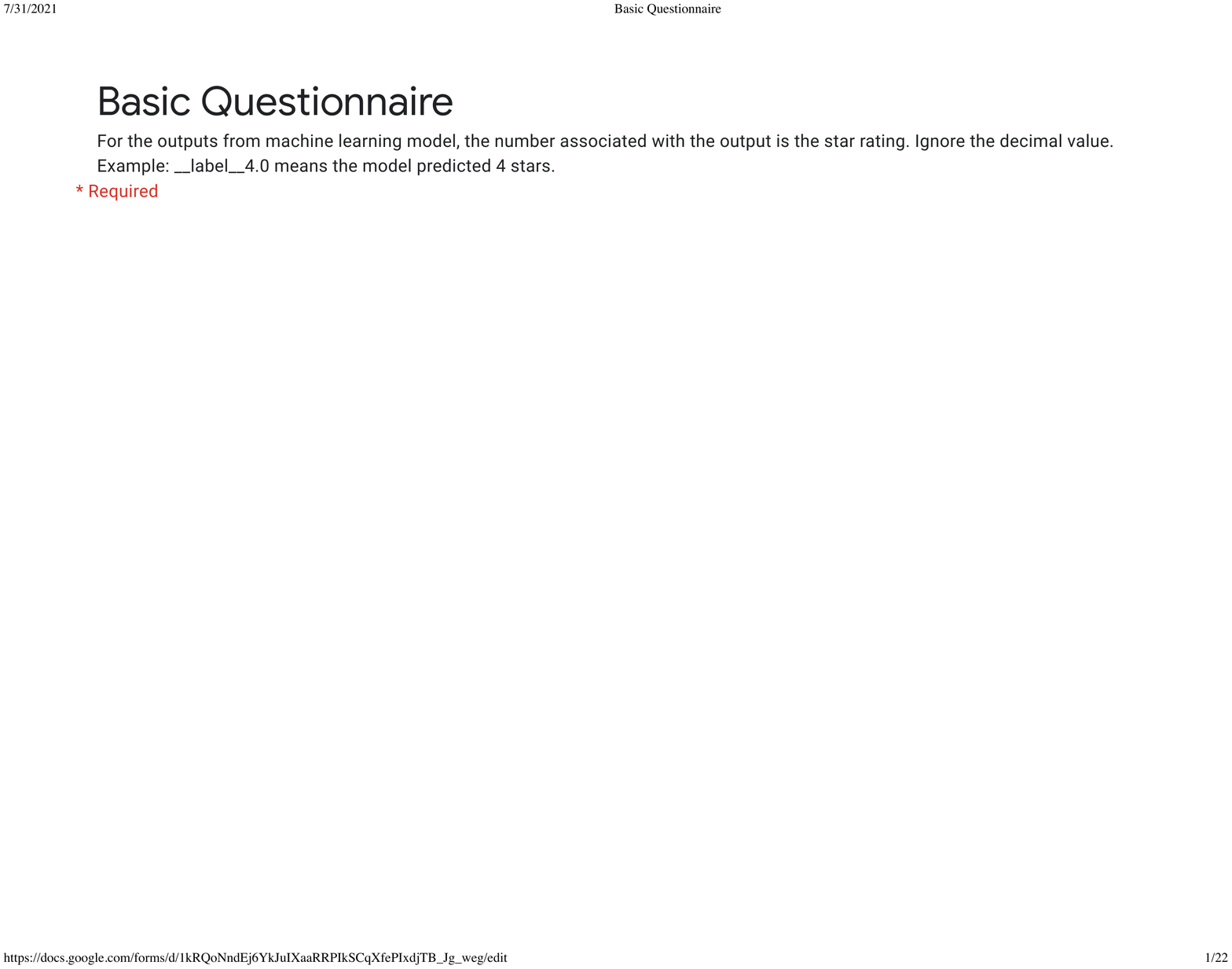}